\documentclass[useAMS,usenatbib]{mn2e}
\usepackage{natbib}
\usepackage{amsmath,amssymb}
\usepackage{graphicx}
\usepackage[dvips]{color}
\usepackage{float}
\usepackage{hyperref}

\bibpunct{(}{)}{;}{a}{}{,} 

\title[Discovery of VHE emission from the blazar 1ES\,033+595]{Discovery of very high energy $\gamma$-ray emission from the blazar 1ES\,0033+595 by the MAGIC telescopes}

\author[J.~Aleksi\'c~et~al.]{
J.~Aleksi\'c$^{1}$,
S.~Ansoldi$^{2}$,
L.~A.~Antonelli$^{3}$,
P.~Antoranz$^{4}$,
A.~Babic$^{5}$,\newauthor 
P.~Bangale$^{6}$,
U.~Barres de Almeida$^{6}$,
J.~A.~Barrio$^{7}$,
J.~Becerra Gonz\'alez$^{8,25}$,\newauthor
% ...
\\
\thanks{Corresponding authors: 
Malwina Uellenbeck, email: malwina.uellenbeck@tu-dortmund.de,  
Nijil Mankuzhiyil, email: mankuzhiyil.nijil@gmail.com,
Saverio Lombardi, email: saverio.lombardi@oa-roma.inaf.it, 
Michele Palatiello, email: michele.palatiello@gmail.com,
Sara Buson, email: sara.buson@pd.infn.it
}(Affiliations can be found after the references)}

\begin{document}

\date{Submitted \today}

\maketitle

%{*} Corresponding authors: Saverio Lombardi (saverio.lombardi@oa-roma.inaf.it), Nijil Mankuzhiyil (mankuzhiyil.nijil@gmail.com), Michele Palatiello (michele.palatiello@gmail.com), Massimo Persic (massimo.persic@gmail.com), Malwina Uellenbeck (malwina.uellenbeck@tu-dortmund.de), and Sara Buson (sara.buson@pd.infn.it)}

%\date{Received: ???~/~Accepted: ???}

\begin {abstract} 
{ 
The number of known very high energy (VHE) blazars is $\sim\,50$, which is very small in comparison to the number of blazars detected in other frequencies. This situation is a handicap for population studies of blazars, which emit about half of their luminosity in the $\gamma$-ray domain. Moreover, VHE blazars, if distant, allow for the study of the environment that the high-energy $\gamma$-rays traverse in their path towards the Earth, like the extragalactic background light (EBL) and the intergalactic magnetic field (IGMF), and hence they have a special interest for the astrophysics community. We present  the first VHE detection of 1ES\,0033+595 with a statistical significance of 5.5\,$\sigma$. The VHE emission of this object is constant throughout the MAGIC observations (2009 August and October), and can be parameterized with a power law with an integral flux above 150 GeV of $(7.1\pm1.3)\times 10^{-12} {\mathrm{ph\,cm^{-2}\,s^{-1}}}$ and a photon index of ($3.8\pm0.7$). 
We model its spectral energy distribution (SED) as the result of inverse Compton scattering of synchrotron photons. For the study of the SED we used simultaneous optical R-band data from the KVA telescope, archival X-ray data by \textit{Swift} as well as \textit{INTEGRAL}, and simultaneous high energy  (HE, $300$\,MeV~--~$10$\,GeV) $\gamma$-ray data from the \textit{Fermi} LAT observatory.  Using the empirical approach of
Prandini et al. (2010) and the \textit{Fermi}-LAT and MAGIC spectra for this object, we estimate the redshift of this source to be $0.34\pm0.08\pm0.05$. This is a relevant result because this source is possibly one of the ten most distant VHE blazars  known to date, and with further (simultaneous) observations could play an important role in blazar population studies, as well as future constraints on the EBL and IGMF.}
\end{abstract}

\begin {keywords}
galaxies: galaxies: BL Lacertae objects: individual (1ES\,0033+595) - gamma rays: galaxies.
\end{keywords}

%\titlerunning{Discovery of 1ES\,0033+595 at very high energy by the MAGIC telescopes}

%\titlerunning{1ES\,0033+595 VHE discovery by MAGIC}

%\authorrunning{Aleksi\'c et~al. (MAGIC collaboration)}

%\maketitle

\section{Introduction}
\label{sec:1}
Blazars are the most commonly detected extragalactic VHE $\gamma$-ray sources, with steadily increasing numbers\footnote{\url{http://tevcat.uchicago.edu/}}  in the very high energy regime (VHE, E\,$>$\,$100$\,GeV) in the past 15 years of ground-based $\gamma$-ray astronomy.
These objects are a subclass of Active Galactic Nuclei (AGNs) with a set of characteristic properties like strong continuum emission extending from the radio all the way to the $\gamma$-ray regime, high polarization (at both optical and radio frequencies) and rapid variability at all frequencies and on all
time scales probed so far. Blazars are thought to be AGNs with jets which are closely aligned with our line-of-sight.
This type of AGN subclass emits a characteristic spectral energy distribution (SED) with at least two broad emission components: one peak with a maximum in optical to X-ray band and a second peak located in the $\gamma$-ray bands \cite{urrypad95}.
The first peak is commonly thought to be related to the synchrotron emission process in magnetic fields of the jets and the second peak is explained as inverse Compton (IC) scattering of low energy photons \cite{res67}. 
If the low energy photons which undergo the IC process are the synchrotron photons, the process is known as the Synchrotron Self Compton (SSC) mechanism  \cite{Tav98}.  
 Alternatively, the origin of the low energy photons can be external to the jet due to external Compton scattering (EC; \citealp{der93}).

1ES\,0033+595 is a blazar near the Galactic plane at cordinates (J2000) R.A.:\,00:35:52.63 and Dec:\,59:50:04.56 \citep{Gio02}, belonging to the BL Lac type. It is classified as an extreme high-frequency peaked (HBL) object with synchrotron emission peaking near $10^{19}$\,Hz  \citep{nie06}.
So far, optical observations of 1ES\,0033+595 were not able to resolve the host galaxy to determine a photometric redshift and thus the redshift of the blazar remains uncertain. A tentative redshift of 0.086 was derived by Perlman, and mentioned in Falomo \& Kotilainen (1999) as a private communication, however to the best of our knowledge the details are not yet published. From the Hubble Space Telescope (HST) images the only information that could be derived was the brightness of the nucleus and an upper limit to the brightness of the surrounding nebulosity, from which a lower limit of $z>0.24$ has been derived  by Sbarufatti et al. (2005). 

 1ES\,0033+595 was first detected as a hard X-ray source by the Einstein Slew Survey in 1992 \citep{elv92}. 
In 1996 it was observed by the HST as part of the snapshot survey of BL Lac objects, and was resolved into two point-like sources with a separation of $1.58~\arcsec$ \citep{sca99}.
These two objects with nearly identical brightness were explained as multiple images of a gravitationally lensed system.
However, the VLA astrometric observations performed in 1997 did not detect a second radio source, ruling out that possibility \citep{rec03}.
 1ES\,0033+595 was observed by the X-ray satellite BeppoSAX in December 1999.
Due to high Galactic absorption it could only be detected in the LECS instrument above $0.4$\,keV and in the PDS instrument up to $\sim$ $60$\,keV \citep{cos01}.
The source was also detected with the \textit{INTEGRAL} satellite in 2003 in the $20$~--~$50$\,keV energy band  with a statistical significance of $5.2$ $\sigma$ \citep{har06}. 
In addition, during  INTEGRAL observations  in 2005 \cite{kui05} the source was a factor of 2.4 brighter.
The fact that all three X-ray observations show different flux levels emphasize very well the variable X-ray nature of this BL Lac object.
With its large field of view and nearly continuous sky coverage the Burst Alert Telescope (BAT) instrument on-board the \textit{Swift} satellite also detected 1ES\,0033+595 during the first 22 months (2004 Dec - 2006 Oct) of observation \citep{tue10}, and has been included in the 22, 58, and 70-month BAT catalogs \footnote{\url{http://swift.gsfc.nasa.gov/results/}}.

In the HE $\gamma$-ray range the source has been continously detected by the \textit{Fermi} Large Area Telescope (LAT). 
The source was reported for the first time after the first 5.5 months of sky survey observations \citep{abd09}.
Since then it has been part of the \textit{Fermi} first bright AGN catalog, with a spectrum consistent with a flat power law with $\Gamma$ = 2.00 $\pm$ 0.13 and a Flux of F($>200$\,MeV) = (20.3 $\pm$ 5.1)$\times 10^{-9} \mathrm{cm^{-2} s^{-1}}$ \citep{abd09}.
In the VHE $\gamma$-ray band this source was first observed for $12$\,h in December 1995 by the Whipple Observatory. 
 These observations yielded only upper limits, F($>350$\,GeV) $<$ $2.1\times 10^{-11} \mathrm{erg\,cm^{-2} s^{-1}}$, equivalent to 20\% of the Crab Nebula flux \citep{hor04}.
 
MAGIC observations of this source were motivated by the BeppoSAX observations \citep{cog02}.
Therein 1ES\,0033+595 was one of the most promising candidate TeV emitters.
MAGIC observed this object in 2006 and later in 2008 for about $5$\,h. From these observations, only a flux upper limit at 95\% confidence level was obtained: F($>170$~GeV) $<2.4\times 10^{-11} \mathrm{cm^{-2} s^{-1}}$, 9.7\%  of the Crab Nebula flux \citep{ale11a}.
New observations in 2009 during the commissioning phase of the MAGIC stereoscopic system led to the discovery of the source in the VHE-$\gamma$-ray range \cite{Mar11,uel12}, as described in this paper. 

\section{Observations and Data Analysis}
\label{sec:obs}
Observations of 1ES\,0033+595 performed in each energy bands are decribed below.
\subsection{VHE data: MAGIC}
\label{sec:MAGICobs}
MAGIC is a system of two $17$\,m-dish Imaging Atmospheric Cherenkov Telescopes (IACTs)
located at the Roque de los Muchachos observatory ($28.8^\circ$N, $17.8^\circ$W, $2200$~m a.s.l.),
in the Canary Island of La Palma. Since 2009 the MAGIC telescopes have carried out stereoscopic
observations with a sensitivity of $<0.8\%$ of the Crab Nebula flux, for energies above $\sim 300$\,GeV
in $50$\,h of observations~\citep{ale12a}. The trigger energy threshold of the system is the lowest among
the currently operating IACTs, with an accessible energy range between $50$\,GeV and several~TeV.

The MAGIC telescopes observed 1ES\,0033+595 from 2009 August 17 until October 14, for a total 
observation time of $23.5$\,h. These observations were performed during the commissioning phase 
of the MAGIC stereoscopic system. During this time, the data taking was carried out using the so-called
``soft stereo trigger mode'', i.e. using the MAGIC-I trigger system and operating MAGIC-II in ``slave mode''. Compared to the standard trigger mode adopted for regular stereoscopic observations
(the so-called ``full stereo trigger mode'', where the Cherenkov events are triggered simultaneously 
by both telescopes), the ``soft stereo trigger mode'' has slightly less sensitivity at the energies 
below $\sim 150$~GeV. In order to take the non-standard trigger condition into account, 
dedicated Monte Carlo (MC) $\gamma$-ray simulations were generated and adopted in this analysis.

The observations of 1ES\,0033+595 were carried out in the so-called wobble mode~\citep{fom94}, 
in which the pointing direction alternates every $20$~min between two positions, offset by $\pm0.4^{\circ}$ 
in R.A. from the source.  The data were taken at zenith angles ranging between $31^{\circ}$ and $35^{\circ}$, which resulted in an analysis energy threshold (defined as the peak of the MC $\gamma$-ray simulated energy distribution for an energy distribution of photon index 3.8 after all analysis cuts) of $\sim90$\,GeV. 

After the application of standard quality checks based on the rate of the stereo events and the 
distributions of basic image parameters, $19.7$\,h of effective on-time events were selected to derive the results.
The rejected data were affected mainly by non-optimal atmospheric conditions during the data taking.

The data analysis was performed using the standard software package MARS~\citep{alb08a,ali09},
including the latest standard routines for the stereoscopic analysis~\citep{lom11,ale12a}.
After the calibration~\citep{alb08b} and the image cleaning of the events,
the information from the individual telescopes is combined and the calculation of basic stereo
image parameters is performed. For the $\gamma$/hadron separation and $\gamma$-direction estimation a 
multivariate method called Random Forest (RF,~\citealp{alb08c}) is applied. For the former task, the 
algorithm employs basic image parameters~\citep{hil85}, timing information~\citep{ali09}, and stereo parameters~\citep{ale12a} 
to compute a $\gamma$/hadron discriminator called \emph{Hadronness} by comparison of real (hadron-dominated) 
data with MC $\gamma$-ray simulations. The \emph{Hadronness} parameter ranges from 0 (for showers confidently 
identified as initiated by $\gamma$-rays) to 1 (for those clearly showing the features of a hadronic 
cosmic-ray initiated shower).
Finally, the estimation of the energy of the events is achieved by averaging individual energy estimators 
for both telescopes based on look-up tables~\citep{ale12a}. 
%For a detail description of the complete MAGIC stereo analysis chain described above see \cite{ale11b}.

The final analysis cuts applied to the 1ES\,0033+595 data were optimized by means of contemporaneous 
Crab Nebula data and MC simulations. In computing the significance of the signal coming from the 1ES\,0033+595 
sky region, single cuts in \emph{Hadronness} and $\theta^{2}$ (see Section~\ref{sec:MAGICresults}) optimized for energies 
close to the threshold were applied\footnote{The cuts correspond to an efficiency for $\gamma$-rays of $>90\%$.}. Conversely, while deriving the spectrum and the light curve of the source, multiple cuts 
optimized in logarithmic energy bins were considered.

\subsection{HE data: \textit{Fermi} LAT}
\label{sec:FERMIobs}
\begin{figure}
\centering
%\vspace{-5cm}
\hspace{-0.5cm}
\includegraphics[width=0.5\textwidth,clip]{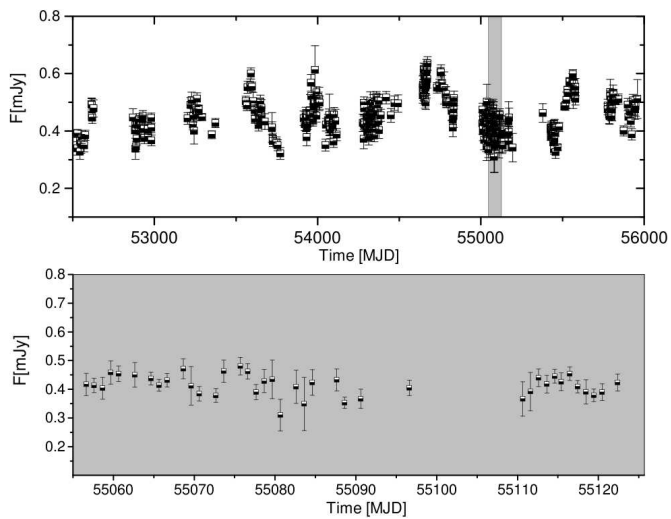}
\caption
{
Optical R-band light curve from 9-year monitoring observations performed by the Tuorla Observatory. The contribution of a nearby star (0.22\,mJy) has not been subtracted.
The MAGIC observation window in 2009 is shown as shaded in the top panel. The bottom panel shows the optical light curve along this time window.
}
\label{fig:0033_opticl_lc}
\end{figure}
The \textit{Fermi} LAT (Large Area Telescope) is a pair conversion telescope designed to cover the energy band from $20$~MeV to more than $300$~GeV. 
Normally it operates in all-sky survey mode, scanning the entire sky every 3\,h. Therefore it can provide observations of 1ES 0033+595 simultaneous with MAGIC. 
\emph{Fermi} data presented in this paper were collected from 2009 August 17 to October 14. 
In addition, to make a comparison with the behavior of the source over a wider time interval, we also show the data analyzed from the beginning of the science phase of the \emph{Fermi} mission, that is 2008 August 4, to 2011 October 28.
Both sets of data were analyzed with the \emph{Fermi} Science Tools package version 09-27-01, available
from the \textit{Fermi} Science Support Center\footnote{http://fermi.gsfc.nasa.gov/ssc/}, and with the post-launch instrument response functions [IRFs, P7SOURCE\_{}V6, \citealp{Ack2012a}].
Only events belonging to the ``Source'' class and located in a circular region of interest (ROI) of $10^{\circ}$ radius, centered at the position of 1ES 0033+595, were selected.
In addition, we excluded photons with zenith angles $>100^{\circ}$ to limit contamination from Earth limb $\gamma$-rays, and data taken when the rocking angle of $Fermi$ was greater than $52^{\circ}$ to avoid time intervals during which the Earth entered the LAT field of view.
The data analysis of 1ES 0033+595 is challenging because this source is located near the Galactic plane.
As a consequence, to reduce the contamination by the Galactic plane diffuse emission, we decided to
restrict the study to the 300\,MeV - 300\,GeV energy range where we can profit from the narrower point-spread function to separate the $\gamma$-ray emission associated with our source from
the intense Galactic foreground.
The analysis in the time interval simultaneous with MAGIC observations was performed using an unbinned maximum likelihood method \citep{1996ApJ...461..396M}.
Instead, a binned maximum likelihood technique was used for the 38 months data set.
%For the 38 month data set instead a binned maximum likelihood technique was used due to the large amount of dataset.
All point sources from the 2FGL catalog \citep{2012ApJS..199...31N} located within $15^{\circ}$ of 1ES 0033+595, and a background model, were included in the model of the region.
The background model used for the analysis includes a Galactic diffuse emission component and an isotropic component (including residual instrumental background), modelled with the files gal\_{}2yearp7v6\_{}v0.fits and iso\_{}P7v6source.txt, which are publicly available\footnote{http://fermi.gsfc.nasa.gov/ssc/data/access/lat/BackgroundModels.html}.
In the full energy range analysis, all point sources within the $10^{\circ}$ radius ROI were fitted with their paramenters set free, while sources beyond $10^{\circ}$ radius ROI had their parameters frozen to the values reported in 2FGL.
The normalizations of the background components were allowed to vary freely.

\subsection{Optical data: KVA}
\label{sec:KVAobs}
The KVA (Kungliga Vetenskapsakademien) telescopes are located at La Palma but operated remotely by the Tuorla Observatory from Finland.
These telescopes are used mainly for optical support observations for the MAGIC telescopes.
Specifically, there is a $60$\,cm telescope which is used for polarimetric observations and a $35$\,cm telescope which performs simultaneous photometric observations with MAGIC. 
Furthermore the smaller $35$\,cm telescope monitors potential VHE $\gamma$-ray candidate AGNs in order to trigger MAGIC observations if one of these selected objects is in a high optical state.
Such observations are performed in the R-band and the magnitude of the source is measured from CCD images using differential photometry. 
During the MAGIC observation of 1ES\,0033+595 the average optical flux obtained by KVA was R=17.93\,magnitude, which corresponds to $0.21$\,mJy. 
To derive this $\nu F_{\nu}$ in the optical band the contribution from a nearby star (0.22\,mJy) was subtracted from the total measured flux (Nilsson et al. 2007).
Moreover, the brightness was corrected for Galactic absorption by R=2.35\,mag (Schlegel et al. 1998). 
The average $\nu F_{\nu}$ during the MAGIC observations corresponds to 8.5$\pm$ 0.5 $\cdot$$\times 10^{-12} \mathrm{erg\,cm^{-2} s^{-1}}$.
As outlined in Figure~\ref{fig:0033_opticl_lc}, during a 9 year KVA survey, the source shows only a marginal flux variability and also during the MAGIC observation 
window (bottom panel in Figure~\ref{fig:0033_opticl_lc}) no significant variability was found.

\begin{figure}
\centering
\vspace{-1.cm}
\hspace{0.0cm}
\includegraphics[width=0.53\textwidth]{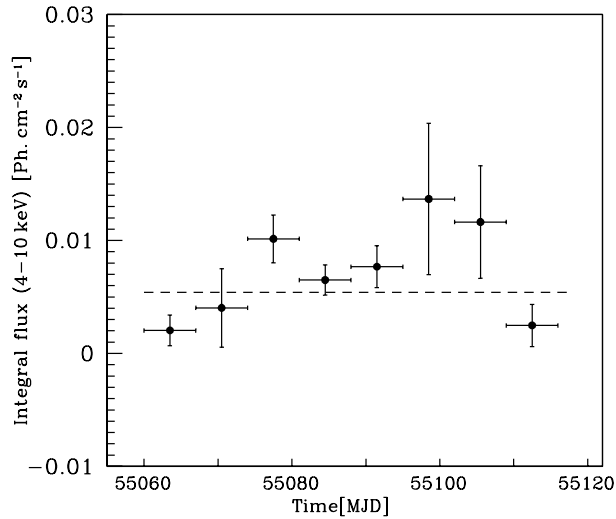}
\vspace{-1.5cm}
\caption
    {The light curve from the MAXI mission in the energy band of 4-10\,keV from 2009 August 17 until October 14 with weekly time binning.  The constant function resulting from the fit to the data is shown as dashed horizontal line.}
\vspace{.5cm}
\label{fig:maxi}
\end{figure}

\begin{figure}
\centering
\vspace{-1.cm}
\hspace{-1.1cm}
\includegraphics[width=0.53\textwidth]{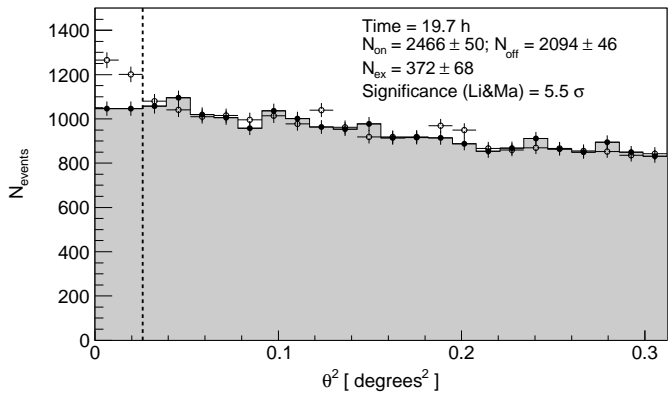}
\caption
{
$\theta^{2}$ distributions of the 1ES\,0033+595 signal (open circle) and background (filled circle) estimation from $19.7$\,h 
of MAGIC stereo observations taken between 2009 August 17 and October 14, above an energy 
threshold of $90$\,GeV. The region between zero and the vertical dashed line (at $0.026$~$\mbox{degrees}^{2}$) 
represents the signal region.
}
\vspace{.5cm}
\label{fig:theta2}
\end{figure}

\begin{figure}
\centering
\vspace{0.cm}
\hspace{-1.2cm}
\includegraphics[width=0.53\textwidth]{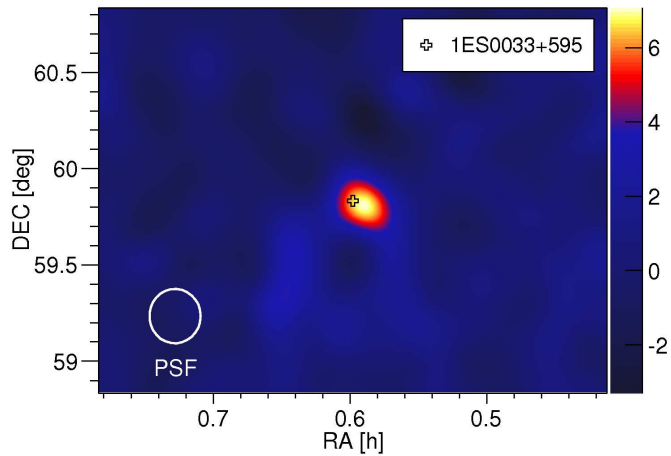}
\caption
{
Significance map of the 1ES\,0033+595 sky region from $19.7$\,h of MAGIC stereo observations above the estimated energy threshold of $90$\,GeV. The color scale represents the test statistic value distribution. The white circle in the lower left indicates the point-spread function (68$\%$ containment) for this analysis.
}
\vspace{.5cm}
\label{fig:sky}
\end{figure}

\subsection{X-ray data}
We extracted a 4-10 keV light curve covering the time frame from 2009 August 17 until October 14 (the MAGIC observation window) from the Monitor of All-sky X-ray Image (MAXI) mission onboard the International Space Station (\citealp{maxi}) \footnote{\url{http://maxi.riken.jp/top/}}.  This light curve is shown in Figure 2. Fitting the X-ray light curve with a constant flux hypothesis yields a flux of $(5.4\pm0.7)\times 10^{-3}$\,photons\,cm$^{-2}$\,s$^{-1}$ with a $\chi^{2}/n_{dof}=19.1/7$, that corresponds to 
a probability $P(\chi^2)=0.008$.
The source shows only marginally significant variability throughout the MAGIC observations.
There are no \textit{Swift}, \textit{XMM} or \textit{INTEGRAL} flux points available which would match the MAGIC observation window.  We have thus used data collected from the ASDC SED Builder tool\footnote{\url{http://tools.asdc.asi.it/SED/}} during the observation period which are closer in time to the MAGIC observation period. All X-ray data points selected for this study are shown in Table\,1.

\begin{table}
\tiny
\caption{The selected X-ray data points in this study.}
% title of Table
%\label{table:1}
% is used to refer this table in the text
\centering
% used for centering table
\setlength{\tabcolsep}{0.40em}
\begin{tabular}{l l l l l}

% centered columns (4 columns)
\hline
% inserts double horizontal lines
\textbf{Mission}& Period &\tiny Frequency & \tiny Flux & \tiny Error  \\
& &\tiny [$10^{18}$\,Hz] & \tiny [$\mathrm{10^{-12}erg\, cm^{-2} s^{-1}}$] & \tiny [$\mathrm{10^{-12}erg\, cm^{-2} s^{-1}}$]  \\
% table heading
\hline
% inserts single horizontal line

\textbf{\textit{XMM}} &July 2004-July 2010& 1.18 &19.5&3.7\\
\textbf{\textit{XMM}} &July 2004-July 2010& 1.53 &14.3&2.1\\
\textbf{\textit{XMM}} &July 2004-July 2010& 0.37 &18.0&2.4\\
\textbf{\textit{Swift}} &Dec 2004-Feb 2008& 9.4 &6.8&0.7\\
\textbf{\textit{Swift}}&Dec 2004-Feb 2008 &  5.1 & 13.7&1.1\\
\textbf{\textit{Swift}}&Dec 2004-Feb 2008&11.5 & 10.7&1.7\\
\textbf{\textit{Swift}}&Dec 2004-Feb 2008 & 6.9 & 15.2&1.0\\
\textbf{\textit{Swift}}&Dec 2004-Sept 2010&7.3 & 8.6&0.8\\
\textbf{\textit{Swift}}&Dec 2004-Sept 2010 & 16.9 & 8.6&0.8\\
\textbf{\textit{INTEGRAL}} &Feb 2003-Apr 2008& 6.8 & 14.2&1.1\\
\textbf{\textit{INTEGRAL}} &Feb 2003-Apr 2008& 15.3 & 9.2&1.0\\
\textbf{\textit{MAXI}} &Aug 2009-Oct 2009& 1.52 & 22.4&2.8\\
%inserts single line
\hline
\end{tabular}
\end{table}

\section{Results}
\label{sec:results}
\subsection{MAGIC}
\label{sec:MAGICresults}

The $\gamma$-ray signal from the source is estimated from the so-called $\theta^{2}$ plot, 
after the application of energy-dependent event selections (including \textit{Hadronness}), and within 
a fiducial $\theta^{2}$ signal region. The parameter $\theta^{2}$ is the squared angular
distance between the reconstructed event direction and the nominal position of the 
expected source. In order to evaluate the residual background of the observation, the $\theta^{2}$ 
distribution around a nominal background control region is also calculated.
Figure~\ref{fig:theta2} shows the $\theta^{2}$ plot for energies above the threshold ($125$\,GeV).
\newline
We found an excess of $372\pm68$ events in the fiducial signal region with $\theta^{2} < 0.026$~$\mbox{degrees}^{2}$,
corresponding to a significance of $5.5~\sigma$, calculated according 
to the Eq.~17 of %~\citet{lima83}.
Li\&Ma (1983).
 Comparing the extension of the excess to the width of the point-spread function of MAGIC (\citealp[$\sim0.1^{\circ}$,][]{ale11b}) we can state that the source has a point-like appearance. Figure~\ref{fig:sky} shows the spatial distribution of the source significance from 1ES\,0033+595. The color scale reports the test statistic value, which is defined as the significance from Li\&Ma (1983) (eq.\,17) applied on a smoothed and modeled background estimation. The fitted position of the signal is RA: 0.588$\pm$0.002 h and DEC: 59.79$\pm$0.02 deg (J2000.0), which, when taking into account the weakness and steepness of the source and the systematic uncertainty in the pointing position of the MAGIC stereo system, is consistent with the catalog cordinates reported in Giommi et al. (2002).
%Moreover, the fitted position of the excess is consistent with the catalog coordinates (Figure~\ref{fig:sky}) given in \citealp{Gio02}. {\bf The color cordinates represents test static value (Li\&Ma 1983 eq. 17, applied on a smoothed and modeled background estimation).}
%Comparing the extension of the excess to the width of the point-spread function of MAGIC (\citealp[$\sim0.1^{\circ}$,][]{ale11b}) 
%($\sim0.1^{\circ}$,~\cite{ale11b}, 
%the source has a point-like appearance. 
\begin{figure}
\centering
%\vspace{-1.cm}
\hspace{-1.cm}
\hspace{-0.cm}
\includegraphics[width=9.cm]{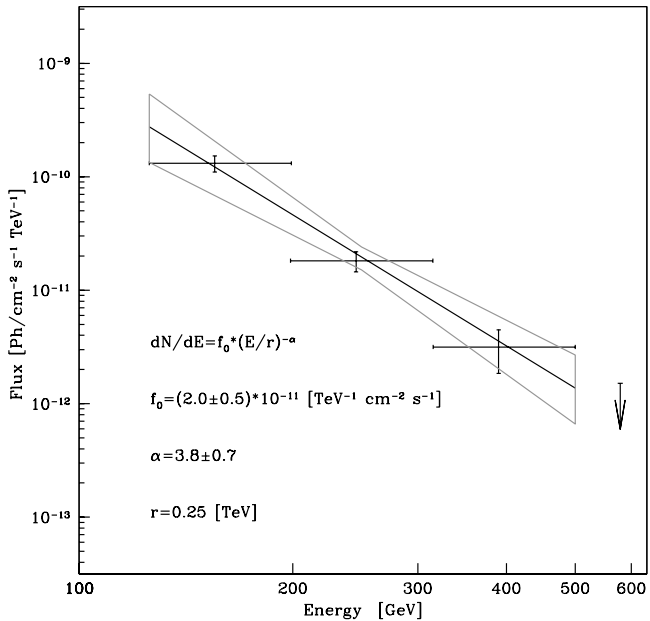}
%\vspace{1.5cm}
\caption{1ES\,0033+595 differential energy spectrum measured by MAGIC between $125$~GeV and $500$~GeV. The power-law fit to the data is shown as a black line, while the uncertainty region is shown as gray bow-tie. A flux upper limit calculated at a 95\% confidence level is also shown in the figure.}
\label{fig:spectrum}
\end{figure}
In Figure~\ref{fig:spectrum}, the unfolded differential energy spectrum of the source derived from the MAGIC 
observations is shown. 
The spectrum between $125$\,GeV and $500$\,GeV can be described by a simple 
power law ($\chi^{2}/n_{dof}=0.45/1$):
\begin{equation}
\frac{\mbox{d}N}{\mbox{d}E} = f_{0} \left(\frac{E}{\mathrm{250~GeV}}\right)^{-\alpha},
\end{equation}
with a photon index of $\alpha=3.8\pm0.7_{stat}\pm0.3_{syst}$\footnote{
The systematic errors of the flux normalization and the photon index considered here have been estimated to be $23\%$ 
and $\pm0.3$, respectively, whereas the systematic error on the energy scale is $17\%$. These values are more conservative 
than those presented in~\cite{ale12a}, given the low flux and the spectral steepness of 1ES\,0033+595, as measured by MAGIC.
},
and a normalization constant at $250$~GeV of $f_{0} =(2.0 \pm 0.5_{stat} \pm 0.5_{syst}) \times 10^{-11} \mathrm{cm^{-2} s^{-1}TeV^{-1}}$.
The mean flux above $150$~GeV is $F_{\gamma}=(7.1\pm1.3_{stat}$~$\pm$~$1.6_{syst})\times10^{-12}~\mathrm{cm^{-2}~s^{-1}}$,
corresponding to $(2.2\pm0.4_{stat}\pm0.5_{syst})\%$~Crab units. Above $\sim$500\,GeV we did not find any gamma-ray excess. We calculated a flux upper limit at a 95\% confidence level using a power-law photon index of 3.8 measured for energies below $\sim$500\,GeV, which yielded $1.52\times10^{-12} \mathrm{ph}$\,$\mathrm{TeV}^{-1}\mathrm{cm}^{-2}\mathrm{s}^{-1}$ at 579 GeV (which is the mean energy when taking into account the energy-dependent detection efficiency of MAGIC and the power-law spectral shape measured below 500 GeV). The obtained upper limit is compatible with the power-law spectrum below 500 GeV.
\newline
This is the first measurement of the differential energy spectrum of 1ES\,0033+595 at VHE $\gamma$-rays.
Furthermore, the energy threshold of MAGIC allows connecting the spectrum to the \textit{Fermi} LAT data points~\citep{abd09}.
%The spectral index measured by MAGIC agrees well with the one measured by \textit{Fermi} LAT above $10$~GeV of $\Gamma=-3.05\pm0.13_{stat}$. 
%%%The emission from 1ES\,0033+595 can be then described by a simple power law between $10$~GeV and $400$~GeV.
Figure~\ref{fig:figure3} shows the weekly time binning light curve of 1ES\,0033+595 data taken by MAGIC between 2009 August 08 and October 10. 
No evidence of variability can be derived from these measurements.
Fitting the light curve with a constant flux hypothesis yields a flux of $7.1\times10^{-12}\,cm^{-2}s^{-1}$ with a $\chi^{2}/n_{dof}=3.7/3$ (which corresponds to 
a probability $P(\chi^2)=0.3$). 
\begin{figure}
\centering
\vspace{-0.5cm}
\includegraphics[width=0.55\textwidth]{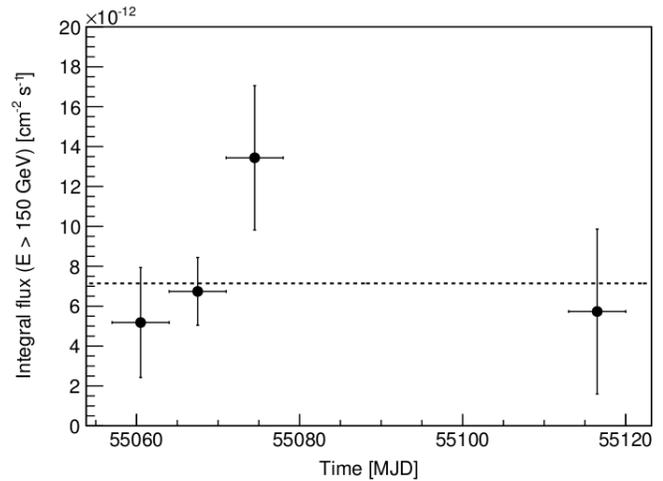}
\caption{1ES\,0033+595 light curve between 2009 August and October above an energy
threshold of $150$\,GeV, and with a weekly time binning. No hints of significant variability
are seen in the data. The dashed horizontal line represents the constant function 
resulting from the fit to the data.}
\label{fig:figure3}
\end{figure}

\subsection{\textit{Fermi} LAT}
\label{sec:FERMIresults}
We obtained the following results for 1ES\,0033+595: for the analysis of data simultaneous with MAGIC observations (2009 August 17 to October 14), the flux above 300\,MeV is ($8.0 \pm 3.6$) $\times 10^{-9}$ cm$^{-2}$ s$^{-1}$ and the photon index is $1.7 \pm 0.2$. 
For the 38 month time interval, the flux above 300\,MeV is ($6.6 \pm 1.0$) $\times 10^{-9}$ cm$^{-2}$ s$^{-1}$ and the photon index is $1.9 \pm 0.1$.
% The light curve obtained during the simultaneous observation time is shown in Figure~\ref{fig:figure3} while the light curve from the whole 38 month period is given in Figure~\ref{fig:fermilc}. 
 The $Fermi$ LAT light curve produced for the whole 38 month period is given in Figure\,7.
For spectral analysis, in the first case we divided the full energy range in 4 energy bins: 2 bins for the 300\,MeV - 10\,GeV range and 2 bins for the 10\,GeV - 300\,GeV range. 
In the latter case we divided the full energy range in 6 logarithmically equal energy bins. 
A separate fit in each energy bin was performed fixing the photon index of all the sources and the isotropic normalization to the values obtained from the likelihood analysis of the full energy range. 
For each energy bin, if the Test Statistic (TS)\footnote{TS is 2 times the difference of the log(likelihood) with and without the source.} value was TS $<9$, then the values of the fluxes were replaced by 2 $\sigma$ confidence level upper limits. 
The latter were computed using the profile method (\citealp{rol05}).

\begin{figure}
\centering
%\vspace{-1.5cm}
\hspace{-1.cm}
\includegraphics[width=9.5cm]{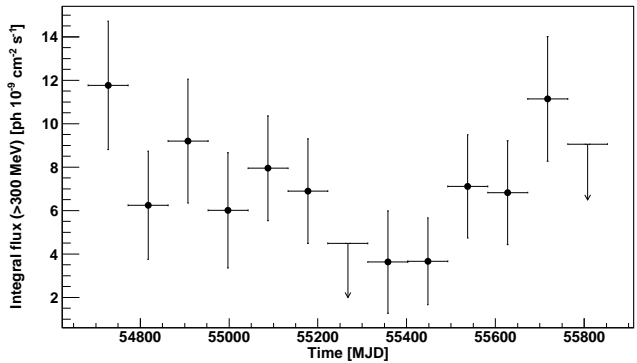}
%\vspace{-5cm}
\caption{The light curve above an energy threshold of 300\,MeV obtained by the \textit{Fermi} LAT during 38 months of observation.}
\label{fig:fermilc}
\end{figure}

Systematic uncertainties in the LAT results of this source were found to be negligible with respect to the statistical ones.
The two major sources of systematic errors that we considered were related to uncertainties in the absolute calibration of the LAT
and to the modeling of interstellar emission. 
The first are due to the uncertainties in the LAT effective area and are typically $\sim$10\% \citep{Ack2012a} and 
therefore negligible compared to the large flux variations observed.
The latter could have had a major impact on the LAT measurements given the vicinity of the source to the Galactic plane.
We decided to investigate them comparing the results obtained using the standard Galactic interstellar model (i.e. gal\_{}2yearp7v6\_{}trim\_{}v0.fits, see footnote\,3) with the results based on eight alternative interstellar emission models.
The eight models were obtained by varying some of the most important parameters of the interstellar emission models,
in a similar way to the approach of Pivato et al. (2013), 
and are based on a subsample of those examined by Ackermann et al. (2012b). 
%As anticipated also these systematic uncertainties 
As expected, the systematic uncertainties due to the modeling of the interstellar emission also were not relevant when compared with the statistical ones. This is due to the  point-source morphology of 1ES 0033+595 and its hard photon spectrum,   both characteristics helping to disentangling the source emission from the Galactic foreground.

\section{Interpretation}

\label{sec:4}

\subsection{Redshift of 1ES\,0033+595 from HE and VHE $\gamma$-ray data}
 \label{sec:4.1}
As already mentioned in Section \ref{sec:1} the redshift of 1ES\,0033+595 is uncertain.
However, for the interpretation of our data (e.g. estimation of the intrinsic VHE $\gamma$-ray spectrum after Extragalactic Background Light -EBL- deabsorption) it is crucial to determine this parameter. 
For this reason we use VHE and HE observations to constrain the redshift of the source by the empirical approach of %\citet{pra11b}.
Prandini et al (2010) (an updated work can also be found at Prandini et al. 2011).
From the findings in Section~\ref{sec:MAGICresults} the VHE spectrum appears to be extremely soft (photon index $\alpha\sim 4$), as would be expected by the absorption of VHE
photons by interaction with the EBL if the source is located at relatively large redshift \cite{Ste92}. 
Such an absorption process depends on the energy of the photon and the distance it has traveled.
The detection of spectra with indices $\Gamma\sim 4$ from blazars located at redshift above 0.2  is consistent with the expectation for EBL absorption\cite{ale12b}.
 
One of the recently developed redshift determination methods is the empirical approach (Prandini et al. 2010), which is based on the assumption that the intrinsic spectrum at TeV energies (e.g. observed by MAGIC) cannot be harder than that in the GeV band (observed by \textit{Fermi} LAT).
%The underlying assumption is, that the VHE spectrum corrected for the absorption of TeV photons by the EBL via photon-photon interaction should still be softer than the $\gamma$-ray spectrum observed by \textit{Fermi} LAT~.
The spectrum shown in Figure~\ref{fig:spectrum} was corrected using an  EBL model from Franceschini et al. (2008) in fine steps of redshift until the slope of the deabsorbed spectrum equals the one in the GeV-band.
In this case a value of $z=0.58\pm0.12$ is obtained which corresponds to an upper limit on the redshift.  An estimate of the likely redshift can be obtained using the inverse formula of Prandini et al. (2010), resulting
%In addition, an estimate of the true distance can be achieved using the linear relation
%between $z^{\ast}$ and the true redshift.
%This relation was the result of a systematic study carried out on blazars with known distances. 
%The extracted redshift of 1ES\,0033+595 using the inverse formula (for more details see~\citealp{pra11b}) results 
 in $z=0.34\pm0.08\pm0.05$,  where the first error (0.08) was calculated based on the uncertainties in photon indices of the source spectra measured by MAGIC and \textit{Fermi} LAT, while the second error (0.05) is the method uncertainty, which comes from the spread in the results after applying the method to known redshift sources.
In the following discussions the new redshift estimation of $z=0.34$ is used.

\subsection{Spectral Energy Distribution}
\label{sec:4.2}
%The SED is reconstructed for the first time from optical to TeV energies for 1ES\,0033+595, allowing us to study the compatibility with a single zone SSC model (Maraschi \& Tavechio 2003) using the $\chi^{2}$-minimization method (Mankuzhiyil et al. 2012).  
%Since SSC emission models generally work well with these kind of sources, EC emission models or hadronic emission models \cite{Mann93} were not considered in this study. 

\begin{figure}
\label{fig:sed}
\centering
\hspace{-1.cm}
\includegraphics[width=9.5cm]{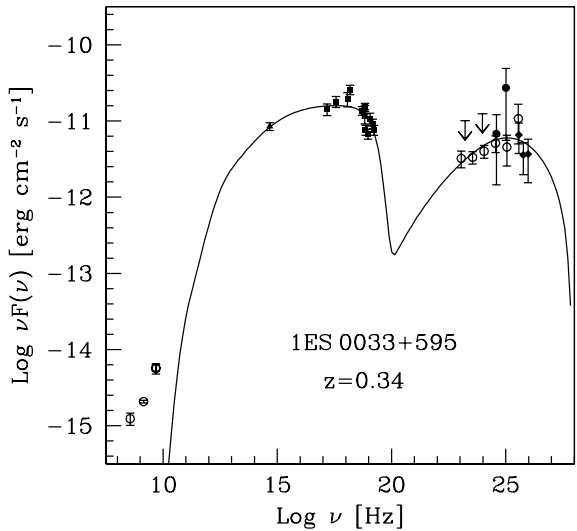}
\vspace{-1cm}
\caption{Broadband SED for 1ES\,0033+595: simultaneous KVA data where the contribution of a nearby star has been subtracted and the flux has been corrected for
Galactic extinction (filled triangle), X-ray data mentioned in Table 1 (filled square), simultaneous \textit{Fermi} LAT data (filled circle) and MAGIC data corrected for the extragalactic
absorption using the model of Franceschini et al. (2008) using a redshift of z=0.34 (filled diamond). We also show the 3 year LAT data (open circle) and archival radio data from the Green Bank and Texas observatory (open circle). The black solid line depicts the one-zone SSC model resulting from the SED model fit described in the text.}
\label{fig:SED}
\end{figure}

The emission characteristics of BL Lac objects are generally well reproduced by the one-zone leptonic model, in
which a population of relativistic electrons inside a region moving down the jet emit through synchrotron and SSC mechanisms \cite{Tav98}. SSC models are generally successful in the modeling of HBLs like 1ES\,0033+595, and will suffice for this first SED modeling of the source. Fitting EC or hadronic \cite{Mann93} models could be attempted in the future as more constraining multi-wavelength data are obtained. 
It can also be noted that the one zone SSC scenario can be a simplified approximation of complex process like emission from an inhomogeneous or stratified region, or a number of independent emission regions. Moreover, as reported in Sikora et al. (1997) the steady state emission can also be parameterized with a number of moving blobs that radiates only while passing through the standing shock. If this is the case, the observer can only see one blob at a given time, which is almost equivalent to the case of single blob emission. With the available data, we cannot distinguish between different scenarios, hence we adopt a single zone emission model (Maraschi \& Tavechio 2003) using the $\chi^{2}$-minimization method fully described
in Mankuzhiyil et al. (2012). The emission region was assumed to be spherical, with radius R, filled with a tangled magnetic field of intensity B and relativistic electrons, emitting synchrotron and SSC radiation. 
The energy distribution of the electrons follows a smoothly-broken power law with normalization K between the Lorentz factors $\gamma_{\rm min}$ and $\gamma_{\rm max}$, with slopes $n_{1}$ and  $n_{2}$ below and above the break at $\gamma_{\rm break}$ . 
The relativistic boosting is represented by the Doppler factor $\delta$.
In Figure~\ref{fig:SED}, we present, for the first time, the reconstructed SED from optical to TeV energies of 1ES\,0033+595. The MAGIC data were corrected for the extragalactic absorption using the model of Franceschini et al. 2008, assuming a redshift of z=0.34.
% where MAGIC data were corrected for the extragalactic absorption using the model of Franceschini et al. (2008), we present for the first time the reconstructed SED from optical to TeV energies of 1ES\,0033+595 using 
%the multiwavelength data as described in Section~\ref{sec:obs}.
Since this source is also very weak in the HE $\gamma$-range, besides the simultaneous LAT spectrum (from 2009 August 17 to October 14), we also included the spectrum from a 3-year time interval (from 2008 August 4 to 2011 October 28) in the SED. The obtained parameter values from the one-zone SSC model fit to the data are summarized in Table\,1.
%we compare in addition the simultaneous \textit{Fermi} LAT spectrum with the behavior of the source over a wider time interval (2008 August 4, to 2011 October 28) in the SED.
From the 3-year LAT analysis a good overlapping between the \textit{Fermi} LAT results and MAGIC is evident.  
The obtained values of the model parameters for the redshift z=0.34, are summarized in Table~\ref{table:2}.
 A comparison with other HBL objects  \citep{Tav10,man11,man12} shows that the one-zone SSC model parameters derived here are compatible with those obtained for other HBL class objects.  We note that, considering the relatively limited experimental constraints, the SSC parameter combination may not be unique. Hence alternative sets of parameters could also provide a satisfactory fit to the data.

\begin{table}
\caption{Parameter values from the one-zone SSC model fit depicted in Figure\,\ref{fig:SED}.}
% title of Table
\label{table:2}
% is used to refer this table in the text
\centering
% used for centering table
\setlength{\tabcolsep}{0.40em}
\begin{tabular}{ l l l l l l r l l l }

% centered columns (4 columns)
\hline
% inserts double horizontal lines
 \tiny $\gamma_{\rm min}$ &\tiny $\gamma_{\rm break}$ &\tiny $\gamma_{\rm max}$ &\tiny $n_{1}$ &\tiny $n_{2}$ &\tiny $B\,[G]$ &\tiny $K\,[\rm cm^{-3}]$ &\tiny $R\,[\rm cm]$&\tiny $\delta$ \\
% table heading
\hline
% inserts single horizontal line
   \tiny 1.0$\cdot  10^{3}$ & \tiny 2.1$\cdot  10^{4}$ & \tiny 2.8$\cdot  10^{6}$ &\tiny 2.0 &\tiny 3.0 &\tiny 1.8$\cdot  10^{-2}$ &\tiny 6.5$\cdot  10^{2}$ &\tiny 8.4$\cdot  10^{16}$& \tiny3.4$\cdot  10^{1}$\\
%inserts single line
\end{tabular}
\end{table}

\section{Summary and Conclusions}
In this paper the first detection of VHE $\gamma$-rays from 1ES\,0033+595 has been reported.
From the 2009 MAGIC data the source is clearly detected at a significance level of $5.5~\sigma$. 
The multiwavelength data presented here confirms the typical HBL blazar subclass behavior of 1ES\,0033+595: 
marginal variability in the optical R-band, a hard spectrum in the \textit{Fermi} LAT regime, and emission of VHE $\gamma$-rays. 
Moreover, the MAGIC detection of 1ES\,0033+595 confirms the identification as a likely VHE $\gamma$-ray emitter by the Costamante \& Ghisellini (2002) list. 
Since the redshift of this source is unknown, but crucial for accurate SED modeling, a new estimation ($z=0.34\pm0.05$) with the empirical approach of Prandini et al. (2010) was performed.
This result is in a good agreement with the lower limit of $z>0.24$ presented in Sbarufatti et al. (2005) and with empirical findings where the sources with redshift greater than 0.2 are characterized 
by a photon index of~$\alpha\sim4$. 
Finally, a comparison with other HBL  objects \citep{Tav10,man11,man12} shows that the model parameters used here for the SED fitting 
are compatible with those obtained for other HBL class objects.
Considering the large  uncertainty in the measured VHE spectrum and the unavailability of simultaneous X-ray data, further work on EBL and IGMF using the presented data is difficult. However,  proper simultaneous MWL coverage on this high redshift object would allow to perform these studies.

\section*{Acknowledgments}
%\begin{acknowledgements}
We would like to thank the Instituto de Astrof\'{\i}sica de
Canarias for the excellent working conditions at the
Observatorio del Roque de los Muchachos in La Palma.
The support of the German BMBF and MPG, the Italian INFN, 
the Swiss National Fund SNF, and the Spanish MICINN is 
gratefully acknowledged. This work was also supported by the CPAN CSD2007-00042 and MultiDark
CSD2009-00064 projects of the Spanish Consolider-Ingenio 2010
programme, by grant DO02-353 of the Bulgarian NSF, by grant 127740 of 
the Academy of Finland,
by the DFG Cluster of Excellence ``Origin and Structure of the 
Universe'', by the DFG Collaborative Research Centers SFB823/C4 and SFB876/C3,
and by the Polish MNiSzW grant 745/N-HESS-MAGIC/2010/0.\\
The \textit{Fermi} LAT Collaboration acknowledges generous ongoing support
from a number of agencies and institutes that have supported both the
development and the operation of the LAT as well as scientific data analysis.
These include the National Aeronautics and Space Administration and the
Department of Energy in the United States, the Commissariat \`a l'Energie Atomique
and the Centre National de la Recherche Scientifique / Institut National de Physique
Nucl\'eaire et de Physique des Particules in France, the Agenzia Spaziale Italiana
and the Istituto Nazionale di Fisica Nucleare in Italy, the Ministry of Education,
Culture, Sports, Science and Technology (MEXT), High Energy Accelerator Research
Organization (KEK) and Japan Aerospace Exploration Agency (JAXA) in Japan, and
the K.~A.~Wallenberg Foundation, the Swedish Research Council and the
Swedish National Space Board in Sweden.
 
Additional support for science analysis during the operations phase is gratefully
acknowledged from the Istituto Nazionale di Astrofisica in Italy and the Centre National d'\'Etudes Spatiales in France.
\\

%--------------

\noindent
$^{1}$ {IFAE, Edifici Cn., Campus UAB, E-08193 Bellaterra, Spain} \\
$^{2}$ {Universit\`a di Udine, and INFN Trieste, I-33100 Udine, Italy} \\
$^{3}$ {INAF National Institute for Astrophysics, I-00136 Rome, Italy} \\
$^{4}$ {Universit\`a  di Siena, and INFN Pisa, I-53100 Siena, Italy} \\
$^{5}$ {Croatian MAGIC Consortium, Rudjer Boskovic Institute, University of Rijeka and University of Split, HR-10000 Zagreb, Croatia} \\
$^{6}$ {Max-Planck-Institut f\"ur Physik, D-80805 M\"unchen, Germany} \\
$^{7}$ {Universidad Complutense, E-28040 Madrid, Spain} \\
$^{8}$ {Inst. de Astrof\'isica de Canarias, E-38200 La Laguna, Tenerife, Spain} \\
$^{9}$ {University of \L\'od\'z, PL-90236 \L\'od\'z, Poland} \\
$^{10}$ {Deutsches Elektronen-Synchrotron (DESY), D-15738 Zeuthen, Germany} \\
$^{11}$ {ETH Zurich, CH-8093 Zurich, Switzerland} \\
$^{12}$ {Universit\"at W\"urzburg, D-97074 W\"urzburg, Germany} \\
$^{13}$ {Centro de Investigaciones Energ\'eticas, Medioambientales y Tecnol\'ogicas, E-28040 Madrid, Spain} \\
$^{14}$ {Technische Universit\"at Dortmund, D-44221 Dortmund, Germany} \\
$^{15}$ {Inst. de Astrof\'isica de Andaluc\'ia (CSIC), E-18080 Granada, Spain} \\
$^{16}$ {Universit\`a di Padova and INFN, I-35131 Padova, Italy} \\
$^{17}$ {Universit\`a dell'Insubria, Como, I-22100 Como, Italy} \\
$^{18}$ {Unitat de F\'isica de les Radiacions, Departament de F\'isica, and CERES-IEEC, Universitat Aut\`onoma de Barcelona, E-08193 Bellaterra, Spain} \\
$^{19}$ {Institut de Ci\`encies de l'Espai (IEEC-CSIC), E-08193 Bellaterra, Spain} \\
$^{20}$ {Japanese MAGIC Consortium, Division of Physics and Astronomy, Kyoto University, Japan} \\
$^{21}$ {Finnish MAGIC Consortium, Tuorla Observatory, University of Turku and Department of Physics, University of Oulu, Finland} \\
$^{22}$ {Inst. for Nucl. Research and Nucl. Energy, BG-1784 Sofia, Bulgaria} \\
$^{23}$ {Universitat de Barcelona (ICC, IEEC-UB), E-08028 Barcelona, Spain} \\
$^{24}$ {Universit\`a di Pisa, and INFN Pisa, I-56126 Pisa, Italy} \\
$^{25}$ {now at NASA Goddard Space Flight Center, Greenbelt, MD 20771, USA and Department of Physics and Department of Astronomy, University of Maryland, College Park, MD 20742, USA}\\
$^{26}$ {now at Department of Physics and Astronomy and the Bartol Research Institute, University of Delaware, Newak, DE 19716,  USA}
$^{27}$ {now at Ecole polytechnique f\'ed\'erale de Lausanne (EPFL), Lausanne, Switzerland} \\
$^{28}$ {now at Department of Physics \& Astronomy, UC Riverside, CA 92521, USA} \\
$^{29}$ {now at Finnish Centre for Astronomy with ESO (FINCA), Turku, Finland} \\
$^{30}$ {now at: Astrophysical Sciences Division, BARC, Mumbai 400085, India}\\
$^{31}$ {also at INAF-Trieste} \\
$^{32}$ {also at Instituto de Fisica Teorica, UAM/CSIC, E-28049 Madrid, Spain} \\
$^{33}$ {now at: Stockholm University, Oskar Klein Centre for Cosmoparticle Physics, SE-106 91 Stockholm, Sweden} \\
$^{34}$ {now at GRAPPA Institute, University of Amsterdam, 1098XH Amsterdam, Netherlands}

\end{document}